# DISCUSSION OF "2004 IMS MEDALLION LECTURE: LOCAL RADEMACHER COMPLEXITIES AND ORACLE INEQUALITIES IN RISK MINIMIZATION" BY V. KOLTCHINSKII

BY STÉPHAN CLÉMENÇON, GÁBOR LUGOSI[1] AND NICOLAS VAYATIS

*Universités Paris 6 & Paris 10, ICREA & Universitat Pompeu Fabra and Université Paris 6*

In this magnificent paper, Professor Koltchinskii offers general and powerful performance bounds for empirical risk minimization, a fundamental principle of statistical learning theory. Since the elegant pioneering work of Vapnik and Chervonenkis in the early 1970s, various such bounds have been known that relate the performance of empirical risk minimizers to combinatorial and geometrical features of the class over which the minimization is performed. This area of research has been a rich source of motivation and a major field of applications of empirical process theory.

The appearance of advanced concentration inequalities in the 1990s, primarily thanks to Talagrand's influential work, provoked major advances in both empirical process theory and statistical learning theory and led to a much deeper understanding of some of the basic phenomena. In the discussed paper Professor Koltchinskii develops a powerful new methodology, *iterative localization*, which, with the help of concentration inequalities, is able to explain most of the recent results and go significantly beyond them in many cases.

The main motivation behind Professor Koltchinskii's paper is based on classical problems of statistical learning theory such as binary classification and regression in which, given a sample $(X_i, Y_i)$, $i = 1, \ldots, n$, of independent and identically distributed pairs of random variables (where the $X_i$ take their values in some feature space $\mathcal{X}$ and the $Y_i$ are, say, real-valued), the goal is to find a function $f : \mathcal{X} \to \mathbb{R}$ whose risk, defined in terms of the expected value of an appropriately chosen loss function, is as small as possible.

In the remaining part of this discussion we point out how the performance bounds of Professor Koltchinskii's paper can be used to study a seemingly

Received March 2006.
[1]Supported by the Spanish Ministry of Science and Technology and FEDER Grant BMF2003-03324 and by the PASCAL Network of Excellence under EC Grant 506778.

different model, motivated by nonparametric *ranking* problems, which has received increasing attention both in the statistical and machine learning literature. Indeed, in several applications, such as the search engine problem or credit risk screening, the goal is to learn how to rank—or to score—observations rather than just classifying them. In this case, performance measures involve pairs of observations, as it can be seen, for instance, with the AUC (Area Under an ROC Curve) criterion. In this setting, empirical risk functionals are no longer simple averages of i.i.d. random variables. Despite this fact, we will see in the sequel that it is possible to apply Professor Koltchinskii's approach. Once again, concentration inequalities will play a major role.

In a ranking problem one has to compare two different observations and decide which one is "better." To formalize the problem in a simple model, let $(X, Y)$ be a pair of random variables taking values in $\mathcal{X} \times \mathbb{R}$ where $\mathcal{X}$ is a measurable space. The random object $X$ models some observation and $Y$ its real-valued label. Let $(X', Y')$ denote a pair of random variables identically distributed with $(X, Y)$, and independent of it. In the ranking problem one observes $X$ and $X'$ but not necessarily their labels $Y$ and $Y'$. We think about $X$ being "better" than $X'$ if $Y > Y'$. The goal is to rank $X$ and $X'$ such that the probability that the better ranked of them has a smaller label is as small as possible. Formally, a *ranking rule* is a function $r : \mathcal{X} \times \mathcal{X} \to \{-1, 1\}$. If $r(x, x') = 1$, then the rule ranks $x$ higher than $x'$. The performance of a ranking rule is measured by the *ranking risk*

$$L(r) = \mathbb{P}\{(Y - Y') \cdot r(X, X') < 0\},$$

that is, the probability that $r$ ranks two randomly drawn instances incorrectly. Clearly, the problem is equivalent to a binary classification problem in which the sign of the random variable $Y - Y'$ is to be guessed based upon the pair of observations $(X, X')$.

Now assume that $n$ independent, identically distributed copies of $(X, Y)$ are available: $D_n = (X_1, Y_1), \ldots, (X_n, Y_n)$. Given a ranking rule $r$, one may use the training data to estimate its risk $L(r) = \mathbb{P}\{(Y - Y') \cdot r(X, X') < 0\}$. The perhaps most natural estimate is the *U-statistic*

$$L_n(r) = \frac{1}{n(n-1)} \sum_{i \neq j} \mathbb{I}_{[(Y_i - Y_j) \cdot r(X_i, X_j) < 0]}.$$

By following the empirical risk minimization strategy studied in the discussed paper, one may consider minimizers of the empirical estimate $L_n(r)$ over a class $\mathcal{R}$ of ranking rules $r : \mathcal{X} \times \mathcal{X} \to \{-1, 1\}$ and study the performance of such empirically selected ranking rules. Define the empirical risk minimizer, over $\mathcal{R}$, by

$$r_n = \arg\min_{r \in \mathcal{R}} L_n(r).$$

In a way analogous to how ordinary empirical risk minimization leads to questions best attacked by using tools of empirical process theory, the key of bounding the performance of an empirical minimizer of the ranking risk is in investigating the properties of $U$-*processes*. For a detailed and modern account of $U$-process theory we refer to the excellent book of de la Peña and Giné [4]. Interestingly, however, bounding the performance of empirical ranking risk minimization boils down to ordinary empirical risk minimization as it is pointed out in the sequel.

We are interested in the risk $L(r_n)$ of the empirical minimizer $r_n$, when compared to $L^*$, the risk of the best possible ranking function $r^*$ in the sense that $L^* = L(r^*) \leq L(r)$ for all measurable ranking functions $r$.

Set first

$$q_r((x,y),(x',y')) = \mathbb{I}_{[(y-y')\cdot r(x,x')<0]} - \mathbb{I}_{[(y-y')\cdot r^*(x,x')<0]}$$

and consider the following estimate of the *excess ranking risk* $\Lambda(r) = L(r) - L^* = \mathbb{E}q_r((X,Y),(X',Y'))$:

$$\Lambda_n(r) = \frac{1}{n(n-1)} \sum_{i \neq j} q_r((X_i,Y_i),(X_j,Y_j)),$$

which is a $U$-statistic of degree 2 with symmetric kernel $q_r$. Clearly, the minimizer $r_n$ of the empirical ranking risk $L_n(r)$ over $\mathcal{R}$ also minimizes the empirical excess risk $\Lambda_n(r)$. To study this minimizer, consider the *Hoeffding decomposition* [5] of $\Lambda_n(r)$:

$$\Lambda_n(r) - \Lambda(r) = 2T_n(r) + W_n(r),$$

where

$$T_n(r) = \frac{1}{n} \sum_{i=1}^{n} h_r(X_i, Y_i)$$

is a sum of i.i.d. random variables with

$$h_r(x,y) = \mathbb{E}q_r((x,y),(X',Y')) - \Lambda(r)$$

and

$$W_n(r) = \frac{1}{n(n-1)} \sum_{i \neq j} \widehat{h}_r((X_i,Y_i),(X_j,Y_j))$$

is a *degenerate* $U$-statistic with symmetric kernel

$$\widehat{h}_r((x,y),(x',y')) = q_r((x,y),(x',y')) - \Lambda(r) - h_r(x,y) - h_r(x',y').$$

The main message of this note is that one can show that, under very general circumstances, the contribution of the degenerate part $W_n(r)$ of the $U$-statistic is negligible compared to that of $T_n(r)$. This means that minimization of $\Lambda_n$ is approximately equivalent to minimizing $T_n(r)$. But since $T_n(r)$ is an average of i.i.d. random variables, this can be studied by techniques worked out for empirical risk minimization, such as the ones in the discussed paper.

The main tool for handling the degenerate part is a new general moment inequality for $U$-processes established in [3], based mostly on concentration and moment inequalities for empirical processes, Rademacher averages, and Rademacher chaos, developed in [1], as well as decoupling and randomization techniques; see [4]. In order to recall this result, we need to introduce some quantities related to the class $\mathcal{R}$. Let $\varepsilon_1, \ldots, \varepsilon_n$ be i.i.d. Rademacher random variables independent of the $(X_i, Y_i)$. Let

$$Z_\varepsilon = \sup_{r \in \mathcal{R}} \left| \sum_{i,j} \varepsilon_i \varepsilon_j \widehat{h}_r((X_i, Y_i), (X_j, Y_j)) \right|,$$

$$U_\varepsilon = \sup_{r \in \mathcal{R}} \sup_{\alpha \colon \|\alpha\|_2 \leq 1} \sum_{i,j} \varepsilon_i \alpha_j \widehat{h}_r((X_i, Y_i), (X_j, Y_j)),$$

$$M = \sup_{r \in \mathcal{R}, k=1,\ldots,n} \left| \sum_{i=1}^n \varepsilon_i \widehat{h}_r((X_i, Y_i), (X_k, Y_k)) \right|.$$

It is shown in [3] that there exists a universal constant $C$ such that, with probability at least $1 - \delta$,

$$\sup_{r \in \mathcal{R}} |W_n(r)| \leq C \left( \frac{\mathbb{E} Z_\varepsilon}{n^2} + \frac{\mathbb{E} U_\varepsilon \sqrt{\log(1/\delta)}}{n^2} + \frac{\mathbb{E} M \log(1/\delta)}{n^2} + \frac{\log(1/\delta)}{n} \right).$$

This inequality bounds the degenerate part of the $U$-process in terms of expected values of certain Rademacher averages and chaoses indexed by $\mathcal{R}$. These quantities have been thoroughly studied and well understood, and may be easily bounded in many interesting cases. For example, it is not difficult to see that if $\mathcal{R}$ is a class of ranking functions of finite VC-dimension $V$, then the right-hand side above is of the order of $(V + \log(1/\delta))/n$.

Whenever $\sup_{r \in \mathcal{R}} |W_n(r)|$ is small, by Hoeffding's decomposition, the minimization of the empirical ranking risk is approximately equivalent to the minimization of the ordinary empirical process $T_n(r)$. This can be studied by the rich theory developed in Professor Koltchinskii's paper. For example, the variance-control techniques mentioned in Section 7 can be applied for this case to derive fast rates of convergence. In particular, it is pointed out in [3] that if there exist constants $c > 0$ and $\alpha \in [0, 1]$ such that for all $r \in \mathcal{R}$,

$$\mathrm{Var}(h_r(X, Y)) \leq c \Lambda(r)^\alpha,$$

DISCUSSION OF LOCAL RADEMACHER COMPLEXITIES 5

then fast rates of convergence may be achieved, depending on the value of $\alpha$ and the modulus of continuity of the empirical process

$$\nu_n(r) = \frac{1}{n}\sum_{i=1}^{n} \ell(r,(X_i,Y_i)) - L(r),$$

where

$$\ell(r,(x,y)) = 2\mathbb{E}\mathbb{I}_{[(y-Y)\cdot r(x,X)<0]} - L(r).$$

For some explicit performance bounds involving these conditions, we refer to [3]. Here we just recall two simple corollaries, derived in [2] in the case when the class $\mathcal{R}$ has a finite VC-dimension $V$:

Assume that

- either: $Y \in \{-1,1\}$ is binary-valued and $\eta(x) = \mathbb{P}\{Y=1|X=x\}$ is such that the random variable $\eta(X)$ has an absolutely continuous distribution on $[0,1]$ with a density bounded by $B$;
- or: $Y = m(X) + \sigma(X)N$ for some (unknown) functions $m:\mathcal{X} \to \mathbb{R}$ and $\sigma:\mathcal{X} \to \mathbb{R}$, and $N$ is a standard normal random variable, independent of $X$ such that $m(X)$ has a bounded density and the conditional variance $\sigma(x)$ is bounded over $\mathcal{X}$.

Then there is a constant $C$ such that for every $\delta, \varepsilon \in (0,1)$, the excess ranking risk of the empirical ranking minimizer $r_n$ satisfies, with probability at least $1-\delta$,

$$L(r_n) - L^* \leq 2\Big(\inf_{r\in\mathcal{R}} L(r) - L^*\Big) + C\varepsilon^{-1}\bigg(\frac{V\log(n/\delta)}{n}\bigg)^{1/(1+\varepsilon)}.$$

Professor Koltchinskii's paper raises many interesting questions about how his new sharp results can be used to prove improved performance bounds for ranking problems. For example, obtaining sharp data-dependent upper confidence bounds so crucial for penalized model selection remains a challenge. We expect that once again, concentration inequalities will be the key for obtaining powerful oracle inequalities for ranking problems.

## REFERENCES

[1] BOUCHERON, S., BOUSQUET, O., LUGOSI, G. and MASSART, P. (2005). Moment inequalities for functions of independent random variables. *Ann. Probab.* **33** 514–560. MR2123200

[2] CLÉMENÇON, S., LUGOSI, G. and VAYATIS, N. (2005). Ranking and scoring using empirical risk minimization. In *COLT* 1–15. MR2203250

[3] CLÉMENÇON, S., LUGOSI, G. and VAYATIS, N. (2006). Ranking and empirical minimization of $U$-statistics. Technical report, Laboratoire de Probabilités et Modèles Aléatoires Université Paris 6.

[4] DE LA PEÑA, V. H. and GINÉ, E. (1999). *Decoupling: From Dependence to Independence.* Springer, New York. MR1666908

[5] HOEFFDING, W. (1948). A class of statistics with asymptotically normal distributions. *Ann. Math. Statist.* **19** 293–325. MR0026294

MODAL'X
UNIVERSITÉ PARIS X – NANTERRE
200, AVENUE DE LA RÉPUBLIQUE
92 000 NANTERRE
FRANCE
E-MAIL: sclemenc@u-paris10.fr

ICREA & DEPARTMENT OF ECONOMICS
UNIVERSITAT POMPEU FABRA
RAMON TRIAS FARGAS 25-27
08005 BARCELONA
SPAIN
E-MAIL: lugosi@upf.edu

LABORATOIRE DE PROBABILITÉS
  ET MODÈLES ALÉATOIRES
UNIVERSITÉ PARIS VI
4, PLACE JUSSIEU
CASE COURRIER 188
75 252 PARIS CEDEX 05
FRANCE
E-MAIL: vayatis@ccr.jussieu.fr